\documentclass[conference]{IEEEtran}

\usepackage{amsthm,amssymb,amsbsy,amsmath,dsfont,latexsym,dsfont,array,layout,graphicx,mathrsfs,color,txfonts,amsfonts,bm,color}

\newtheorem{definition}{Definition}
\newtheorem{proposition}[definition]{Proposition}
\newtheorem{lemma}[definition]{Lemma}

\newtheorem{theorem}[definition]{Theorem}

\newtheorem{example}[definition]{Example}

\begin{document}

\abovedisplayskip0.85\abovedisplayskip

\sloppy

\title{Stabilizer Formalism for Generalized Concatenated Quantum Codes}


 \author{
   \IEEEauthorblockN{
     Yun-Jiang Wang\IEEEauthorrefmark{1}\IEEEauthorrefmark{2},
     Bei Zeng\IEEEauthorrefmark{3},
     Markus Grassl\IEEEauthorrefmark{4}
      and
     Barry C. Sanders\IEEEauthorrefmark{1}}

   \IEEEauthorblockA{
     \IEEEauthorrefmark{1}Institute for Quantum Science and Technology,
      University of Calgary,\\
      Calgary, Alberta T2N 1N4, Canada, Email: sandersb@ucalgary.ca
     }
   \IEEEauthorblockA{
     \IEEEauthorrefmark{2}State Key Laboratory of Integrated Services Networks,
     Xidian University,\\ Xi'an, Shaanxi 710071, China
     Email: yunjiang.w@gmail.com}
   \IEEEauthorblockA{
     \IEEEauthorrefmark{3}Department of Mathematics and Statistics,
      University of Guelph,
      \\Guelph, Ontario N1G 2W1, Canada, Email: zengb@uoguelph.ca}
   \IEEEauthorblockA{
     \IEEEauthorrefmark{4}Centre for Quantum Technologies, National University of Singapore,\\
      3 Science Drive 2, Singapore 117543, Singapore, Email: Markus.Grassl@nus.edu.sg}
}



\maketitle

\begin{abstract}
The concept of generalized concatenated quantum codes (GCQC) provides a
systematic way for constructing good quantum codes from short
component codes. We introduce a stabilizer formalism for GCQCs, which
is achieved by defining quantum coset codes. This formalism offers a
new perspective for GCQCs and enables us to derive a lower bound on
the code distance of stabilizer GCQCs from component codes parameters,
for both non-degenerate and degenerate component codes. Our formalism
also shows how to exploit the error-correcting capacity of component
codes to design good GCQCs efficiently.
\end{abstract}

\section{Introduction}

Error-correcting codes are necessary to overcome restrictions in
computation and communication due to noise, but developing algorithms
for finding `good' codes is generically an intractable problem and
evidently the central question of coding theory. `Good codes' are
special in that they have good trade-off among rate, distance,
encoding and decoding costs, thereby reducing requisite space and time
resources.

In classical settings, constructing generalized concatenated codes,
which incorporate multiple outer codes concatenated with multiple
inner codes, is a promising approach for realizing good trade-off
among those parameters~\cite{LD04, D98}. Recently, generalized
concatenation has been introduced into the quantum scenario, providing
a systematic way to construct good quantum codes with short component
codes~\cite{GSSSZ09, GSZ09}.

The stabilizer formalism plays a central role in almost all branches
of quantum information science, especially in quantum coding
theory. Stabilizer codes, which are quantum analogues of classical
linear codes, form the most important class of quantum
error-correcting codes (QECCs)~\cite{Got97, CRSS98}. The stabilizer
formalism serves not only a role analogous to the classical
parity-check matrix, but also takes a role analogous to the classical
generator matrix during the decoding and encoding
procedures~\cite{Got97,NC00}.  However, the stabilizer formalism for
generalized concatenated quantum codes (GCQCs) has not been
investigated in much detail previously, and the understanding of GCQCs
is still far from satisfactory compared to their classical
counterparts.

In this work we introduce the stabilizer formalism for GCQCs, thereby
providing a new perspective for the GCQC framework as well as a
powerful and systematic technique for constructing good stabilizer
codes. By using our stabilizer formalism, we derive a lower bound on
the achievable distance for GCQCs. Moreover, our stabilizer formalism
for GCQCs clarifies how to exploit the error-correcting capacity of
component codes to improve the performance of the resultant codes
efficiently.

\section{Generalized concatenated stabilizer codes}


A qudit is a quantum system modeled by a $q$-dimensional Hilbert space
$\mathbb{C}^q$, where $q$ is a prime power. A stabilizer (or additive)
quantum code encoding $k$ qudits into an $n$-qudit system, with
minimum distance $d$, is denoted by $[\![n,k,d]\!]_q$.

\subsection{Idea of generalized concatenated quantum codes}
A concatenated stabilizer code is constructed from two component
quantum codes: an \emph{outer} code $A$ with parameters
$[\![N,K,D]\!]_Q$ and an \emph{inner} code $B$ with parameters
$[\![n,k,d]\!]_q$,\footnote{In the sequel, we usually denote the outer
  parameters by \emph{capital} Latin characters and the inner
  parameters by their \emph{small} counterparts.} such that
$Q=q^{k}$. The concatenated code $A\circ B$ is constructed in the
following way: for any state $|\phi\rangle=\sum_{j_1\cdots
  j_N}\alpha_{j_1\cdots j_N}|{j_1\cdots j_N}\rangle$ of the outer code
$A$, replace each basis vector $|j_l\rangle$ (where $j_l=0,\cdots,Q-1$
for $l=1,\dots,N$) by a basis vector $|\psi_{j_l}\rangle$ of the inner
code $B$. This mapping yields

\begin{equation}\label{eq:basisofcgc}
|\phi\rangle \mapsto |\widetilde{\phi}\rangle=\sum_{j_1\cdots j_N}\alpha_{j_1\cdots j_N}|\psi_{j_1}\rangle\cdots |\psi_{j_N}\rangle,
\end{equation}
and the resultant code is an $[\![nN,kK,\mathcal{D}]\!]_q$ stabilizer code where $\mathcal{D}\geq dD$~\cite{Got97,KL96}.

For GCQCs, the role of the basis vectors of the inner quantum code is taken on by \emph{subcodes} of the inner code~\cite{GSSSZ09}.
In its simplest version (two-level version), a GCQC is also constructed from two quantum codes: an outer code $A_1$ with with parameters $[\![N,K_1,D_1]\!]_{Q_1}$ and an inner code $B_1$ with parameters $[\![n,k_1,d_1]\!]_q$, such that the inner code $B_1$ could be further partitioned into $Q_1$ subcodes $\{B_2^{(j)}\}^{Q_1-1}_{j=0}$, i.\,e.,
\begin{equation}\label{partition1}
B_1=\bigoplus_{j=0}^{Q_1-1} B_2^{(j)}.
\end{equation}
and each $B_2^{(j)}$ is an $[\![n,k_2,d_2]\!]_q$ code, with basis vectors $\{|\psi_{i}^{(j)}\rangle\}^{q^{k_2}-1}_{i=0}$, $j=0,\dots,Q_1-1$ and $d_2\ge d_1$. Thus we have $q^{k_1-k_2}=Q_1$.\footnote{The resultant code is reduced to the usual concatenated stabilizer code when $k_2=0$.}

To construct a GCQC, replace each basis state $|j\rangle$ of the outer
code $A_1$ with a basis state $\{|\psi_{i}^{(j)}\rangle\}$ of
$B_2^{(j)}$. In this way, each basis state $|j\rangle$ of the outer
code is mapped to the subcode $B_2^{(j)}$. Consequently, given a state
$|\phi\rangle=\sum_{j_1\cdots i_N}\alpha_{j_1\cdots j_N}|j_1\cdots
j_N\rangle$ of the outer code together with an unencoded basis state
$|i_1 \cdots i_N\rangle\in({\mathbb{C}}^{{q}^{k_2}})^{\otimes N}$, the
encoding of a GCQC is given by the following mapping~\cite{GSSSZ09}:
\begin{equation}\label{basisofGQC}
|\phi\rangle|i_1 \cdots i_N\rangle \mapsto \sum_{j_1\cdots j_N} \alpha_{j_1\cdots j_N}|\psi_{i_1}^{(j_1)}\rangle\cdots |\psi_{i_N}^{(j_N)}\rangle.
\end{equation}

This then gives a GCQC code with parameters
$[\![\mathcal{N},\mathcal{K},\mathcal{D}]\!]_q$, where
$\mathcal{N}=nN$, $\mathcal{K}=(k_1-k_2)K_1+k_2N$, and the minimum
distance $\mathcal{D}$ to be determined. Note that the basis states
$|i_1 \cdots i_N\rangle$ span a trivial outer code
$[\![N,N,1]\!]_{Q_2}$, where $Q_2=q^{k_2}$. Therefore, two outer codes
and two inner codes are used, which is where the name `two-level
concatenation' comes from.

\subsection{Quantum coset codes}
We adapt the concept of coset codes~\cite{LD04,ForneyI,ForneyII} to the quantum scenario to provide an alternative understanding for
stabilizer GCQCs. Coset codes will help to build a systematic interpretation for GCQCs from the viewpoint of the stabilizer formalism.


We choose any subcode $B_2^{(j)}$ in the decomposition (\ref{partition1}) and denote it as $B_2$.
Continuing the partitioning process, say
\begin{equation}\label{partitionm}
B_i=\bigoplus_{j=0}^{Q_i-1} B_{i+1}^{(j)},
\end{equation}
for $i=2,3,\ldots,m$,
we obtain a chain of subcodes
\begin{equation}\label{subsetsrelation}
B_{m+1}\subset B_{m}\subset \cdots \subset B_3\subset B_2\subset B_1,
\end{equation}
where all subcodes $B_i^{(j)}$ on level $i$ have parameters
$[\![n,k_i,d_i]\!]_q$.  To simplify notation, we use $B_i$ to denote
any of the subcodes $B_i^{(j)}$.  On level $m+1$, all subcodes are
one-dimensional subspaces, and we choose
$B_{m+1}=\{|{\mathbf{0}}\rangle\}$.

As the subspaces $B^{(j)}_{i+1}$ in the decomposition
(\ref{partitionm}) are all isomorphic, we can, on an abstract level,
rewrite the decomposition as a tensor product of a vector space of
dimension $Q_i$, spanned by orthonormal states $|j\rangle$
corresponding to the indices $j$ in the decomposition
(\ref{partitionm}), and the subcode $B_{i+1}$.  We denote this
situation by
\begin{equation}
\label{codewordofinnercodes2}
B_{i}= [\![B_{i}/B_{i+1}]\!]\otimes B_{i+1}.
\end{equation}
It turns out that the co-factor $[\![B_{i}/B_{i+1}]\!]$ in
(\ref{codewordofinnercodes2}) can be identified with an additive
quantum code of dimension
\begin{equation}\label{dimension}
Q_i = \dim [\![B_{i}/B_{i+1}]\!]=q^{k_{i}-k_{i+1}}.
\end{equation}
Note that both $B_{i+1}$ and $[\![B_i/B_{i+1}]\!]$ are defined with
respect to a quantum system with $n$ qudits.  In analogy to coset
codes in the context of generalized concatenated codes
\cite{LD04,ForneyI,ForneyII}, we call $[\![B_{i}/B_{i+1}]\!]$ a
\emph{quantum coset code}.

This then directly leads to
\begin{equation}\label{partitionofinnercodes}
B_1=[\![B_1/B_2]\!]\otimes[\![B_2/B_3]\!]\otimes \dots \otimes B_{m},
\end{equation}
i.\,e., the quantum code $B_1$ is abstractly a tensor product of $m$
coset codes $[\![B_1/B_2]\!]$, $[\![B_2/B_3]\!]$,\ldots,
$[\![B_{m}/B_{m+1}]\!]=B_{m}$. These $m$ quantum coset codes will be
used as inner codes to be concatenated with $m$ outer codes $A_i$
($i=1,2,\ldots m$) to form an $m$-level concatenated quantum code.

On each level, the basis state $|j\rangle\in{\mathbb{C}}^{Q_i}$ of the
`coordinate space' of the outer code $A_i=[\![N_i,K_i,D_i]\!]_{Q_i}$
is mapped to the basis index $j$ of the corresponding quantum coset
code $[\![B_i/B_{i+1}]\!]$.  Hence, the $i$th level of concatenation yields
the concatenated code
\begin{equation}
\label{ithlevelcodeword}
C_i=A_i\circ [\![B_{i}/B_{i+1}]\!].
\end{equation}
The resultant $m$-level concatenated code $C$ is then the abstract
tensor product of those $m$ concatenated codes, i.\,e.,
\begin{equation}
\label{finalcodeword}
C=C_1 \otimes C_2 \otimes \cdots \otimes C_{m}.
\end{equation}

\section{Stabilizer formalism for generalized concatenated quantum codes}

\subsection{Stabilizers for the inner codes}

We now develop the stabilizer formalism for GCQCs based on the coset codes $[\![B_{i}/B_{i+1}]\!]$.
For simplicity we consider the case $q=2$, i.\,e., all codes $B_i$s are qubit stabilizer codes. The extension to larger dimensions $q$ is straightforward.

For the code $B_1=[\![n,k_1,d_1]\!]_2$, let
$S_{B_1}=\left\{g_1,g_2,\ldots,g_{n-k_1}\right\}$ denote the set of generators of
the stabilizer group.  The corresponding sets of logical $X$- and
$Z$-operators for the $k_1$ encoded qubits are denoted by
$\overline{X}_{B_1}=\left\{\overline{X}_{1},\overline{X}_{2},\dots,\overline{X}_{k_1}\right\}$
and
$\overline{Z}_{B_1}=\left\{\overline{Z}_{1},\overline{Z}_{2},\dots,\overline{Z}_{k_1}\right\}$.
Similarly, for the code $B_i$, we use $S_{B_i}$, $\overline{X}_{B_i}$,
and $\overline{Z}_{B_i}$ to denote the set of stabilizer generators,
the logical $X$-, and the logical $Z$-operators, respectively.

Note that $B_i=[\![n,k_i,d_i]\!]_q$ is a subcode of $B_1$, for $2\le
i\le m+1$. Thus $S_{B_i}$ can be chosen as the union of $S_{B_1}$ and
a set comprising $k_1-k_i$ commuting logical operators of $B_1$,
which is denoted as $\hat{S}_{B_i}$.  Without loss of generality, we
choose
$\hat{S}_{B_i}=\left\{\overline{Z}_{1},\overline{Z}_{2},\dots,\overline{Z}_{k_1-k_i}\right\}$. Thus
we have
\begin{IEEEeqnarray}{rcl}
S_{B_i}=S_{B_1} \cup \hat{S}_{B_i}&{}={}&\left\{g_1,g_2,\dots,g_{n-k_1},\overline{Z}_{1},\dots,\overline{Z}_{k_1-k_i}\right\},
  \label{eq:sbi}\\
\overline{Z}_{B_i}&=&\left\{\overline{Z}_{k_1-k_i+1},\overline{Z}_{k_1-k_i+2},\dots,\overline{Z}_{k_1}\right\},
  \label{eq:logicalzi}\\
\overline{X}_{B_i}&=&\left\{\overline{X}_{k_1-k_i+1},\overline{X}_{k_1-k_i+2},\dots,\overline{X}_{k_1}\right\}.
  \label{eq:logicalxi}
\end{IEEEeqnarray}

Note that eventually we will arrive at $B_{m+1}=\{|\overline{00 \cdots
0}\rangle\}$. This logical state $|{\mathbf 0}\rangle$ is the only
vector shared by all $B_i$.

\begin{figure}[htbp]
\vskip-5mm
\centerline{\includegraphics[width=9cm]{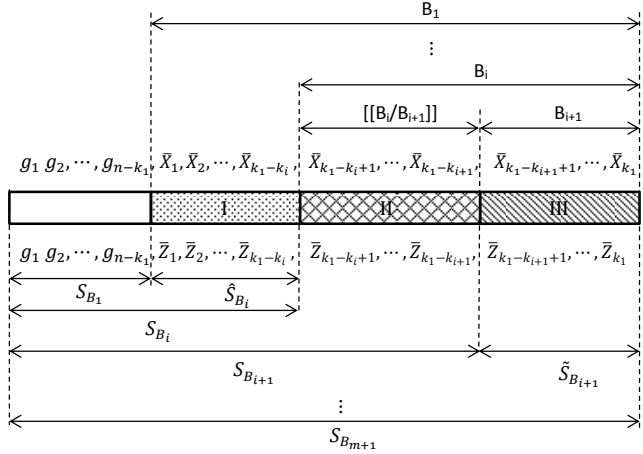}}
\vskip-5mm
\caption{Structure of a quantum coset code obtained by nesting $B_i$s
  in computational basis. Here I, II, and III indicate the code
  factors $[\![B_1/B_i]\!]$, $[\![B_i/B_{i+1}]\!]$, and $B_{i+1}$,
  respectively.  They are spanned by the states obtained when the
  corresponding logical operators located in their area act on the
  logical state $|\mathbf{0}\rangle$ shared by all subcodes
  $B_i$. All other terms are defined in the main text.}
 \label{fig:nesting}\vskip-2mm
\end{figure}

Recall that the code $[\![B_{i}/B_{i+1}]\!]$ is an additive quantum code
with dimension $Q_i=q^{k_{i}-k_{i+1}}$.  Let $S_{[\![B_{i}/B_{i+1}]\!]}$
denote the set of generators of its stabilizer group.  Defining the
set
$\tilde{S}_{B_{i+1}}=\overline{Z}_{B_{i+1}}=\{\overline{Z}_{k_1-k_{i+1}+1},\overline{Z}_{k_1-k_{i+1}+2},\dots,\overline{Z}_{k_1}\}$,
we have
\begin{IEEEeqnarray}{rcl}
S_{[\![B_{i}/B_{i+1}]\!]}&{}={}&S_{B_{i}}\cup\tilde{S}_{B_{i+1}}\nonumber\\
&=&\left\{g_1,\dots,g_{n-k_1},\;
  \overline{Z}_1,\dots,\overline{Z}_{k_1-k_{i}},\;
  \overline{Z}_{k_1-k_{i+1}+1},\dots,\overline{Z}_{k_1}\right\}.\nonumber\\
\end{IEEEeqnarray}
The logical operators of $[\![B_i/B_{i+1}]\!]$ are
\begin{IEEEeqnarray}{rcl}
\overline{Z}_{[\![B_{i}/B_{i+1}]\!]}&{}={}&
  \left\{\overline{Z}_{k_1-k_{i}+1},\overline{Z}_{k_1-k_{i}+2},\dots,\overline{Z}_{k_1-k_{i+1}}\right\}\label{logicalZofcosetcode}\\
\llap{\text{and}\quad}\overline{X}_{[\![B_{i}/B_{i+1}]\!]}&=&
 \left\{\overline{X}_{k_1-k_{i}+1},\overline{X}_{k_1-k_{i}+2},\dots,\overline{X}_{k_1-k_{i+1}}\right\}.\label{logicalXofcosetcode}
\end{IEEEeqnarray}
The structure is illustrated in Fig.~\ref{fig:nesting}.

\subsection{Stabilizers for the generalized concatenated quantum codes}



We now discuss the stabilizers for a GCQCs with an inner code $B_1$
and its $m$-level partitions as given in
Eq.~(\ref{subsetsrelation}). We will have $m$ outer codes $A_i$,
$i=1,2,\ldots,m$, each with parameters $[\![N,K_i,D_i]\!]_{Q_i}$.

We first account for the stabilizer generators obtained solely from
$B_1$.  This set is denoted by $S_I$.  The resulting GCQC has length
$nN$. For each sub-block  of length $n$, we have stabilizer generators
from $S_{B_1}$ acting on that block. We can express $S_I$ as
\begin{IEEEeqnarray*}{rcccccccccccccccccccccccc}
{S_I}&{}={}&
        S_{B _1} &{}\otimes{}&\{\rm id\}&{}\otimes{}&\{\rm id\}&{}\otimes{}&\cdots&{}{}\otimes{}&\{\rm id\}\\
&\cup&\{\rm id\}&{}\otimes{}& S_{B_1} &{}\otimes{}&\{\rm id\}&{}\otimes{}&\cdots&{}{}\otimes{}&\{\rm id\}\\
&    &\vdots    && \vdots && \vdots && \ddots && \vdots\\
&\cup&\{\rm id\}&{}\otimes{}&\{\rm id\}&{}\otimes{}&\{\rm id\}&{}\otimes{}&\cdots&{}\otimes{}& S_{B_1},
\end{IEEEeqnarray*}
where ${\rm id}$ denotes the identity operator on $n$ qubits, and the
tensor product of two sets is defined as $S\otimes T=\{s\otimes
t\colon s\in S, t\in T\}$. Evidently, there are in total
$(n-k_1)\times N$ independent generators in $S_I$.

Next we consider the contributions from the outer codes
$A_i=[\![N,K_i,D_i]\!]_{Q_i}$.  For $q=2$, we have $Q_i=2^{r_i}$,
where $r_i=k_{i}-k_{i+1}$.  Each $A_i$ is a subspace of
$(\mathbb{C}^{Q_i})^{\otimes N}\cong(\mathbb{C}^{2})^{\otimes
  r_iN}$. Hence each $A_i$ can be viewed as a subspace of $r_iN$
qubits which can be grouped into $N$ blocks with $r_i$ qubits in each
block.  Denote the set of generators for the stabilizer of $A_i$ by
$S_{A_i}$. Any operator $G\in S_{A_i}$ can be expressed as
$G=\bigotimes_{j=1}^N G_j$, where each $G_j=X^{\alpha}\cdot
Z^{\beta}$, $\alpha,\beta\in GF(2^{r_i})$ is a generalized Pauli
operator on $\mathbb{C}^{2^{r_i}}$, which can be further represented as
$(X^{a_1}X^{a_2}\cdots X^{a_{r_i}})\cdot (Z^{b_1}Z^{b_2}\cdots
Z^{b_{r_i}})$, where $a_j,b_j\in GF(2)$ for
$j={1,2,\dots,r_i}$~\cite{AK2001}.\footnote{Here we omit the tensor
  product symbol, i.\,e., $X^{a_1}X^{a_2}\cdots
  X^{a_{r_i}}$ is to be read as $X^{a_1}\otimes X^{a_2}\otimes \cdots \otimes
  X^{a_{r_i}}$, similarly for $Z^{b_1}Z^{b_2}\cdots Z^{b_{r_i}}$.}


Note that each $X^{a_{\ell}}$ ($Z^{b_{\ell}}$) is a Pauli operator
corresponding to the $\ell$th qubit for each block with $r_i$
qubits. For the concatenation at the $i$th level, each basis vector
$|j\rangle$ of the `coordinate space' of $A_i$ will be mapped to a
basis vector $|b_j^{(i)}\rangle$ of the coset code
$[\![B_{i}/B_{i+1}]\!]$. Therefore, in order to import the constraints
coming from the stabilizer generators $S_{A_i}$, we need to replace
the Pauli operators $X^{a_l}$ ($Z^{b_l}$) for each block of $r_i$
qubits by the corresponding logical operators of $[\![B_{i}/B_{i+1}]\!]$,
which are given by Eqs.~(\ref{logicalZofcosetcode}) and
(\ref{logicalXofcosetcode}).

For each of the $N$ blocks in total, this procedure encodes $r_i$
qubits into $n$ qubits. For each $G_j=X^{\alpha}\cdot Z^{\beta}$,
($1\le j\le N$), the replacement mentioned above yields
\begin{IEEEeqnarray}{rcl}
\overline{G_j}&{}={}&
(\overline{X}_{k_1-k_i+1}^{a_1}\overline{X}_{k_1-k_i+2}^{a_2}\cdots\overline{X}_{k_1-k_{i+1}}^{a_{r_i}})\nonumber\\
&&\quad{}\cdot
(\overline{Z}_{k_1-k_i+1}^{b_1}\overline{Z}_{k_1-k_i+2}^{b_2}\cdots\overline{Z}_{k_1-k_{i+1}}^{b_{r_i}}).
\end{IEEEeqnarray}
Thus each generator $G\in S_{A_i}$ is mapped to
$\overline{G}=\bigotimes_{j=1}^N \overline{G}_j\in
\overline{S}_{A_i}$, where $\overline{S}_{A_i}$ denotes the resulting
set of generators after the replacement.

For each outer code $A_i$, denote the set of logical operators by
$L_{A_i}$.  Then using a similar replacement as for the stabilizer
generators, we obtain a set of logical operators for the $i$th level
concatenated code, which we denote by $\overline{L}_i$.  We then have the
following proposition, which is a direct consequence of
Eq.~(\ref{finalcodeword}).
\begin{proposition}
\label{pro:stabilizer}
The set of stabilizer generator $S_{\mathcal{C}}$ for the generalized
concatenated quantum code
$\mathcal{C}=[\![\mathcal{N},\mathcal{K},\mathcal{D}]\!]_q$ is given
by
\begin{equation}
S_{\mathcal C}=S_I\cup\bigcup_{i=1}^m \overline{S}_{A_i},
\end{equation}
and the set of logical operators $L_{\mathcal C}$ is given by
\begin{equation}
\label{eq:logicalC}
L_{\mathcal C}=\bigcup_{i=1}^m \overline{L}_i.
\end{equation}
Note that we may multiply any logical operator by an element of the
stabilizer without changing its effect on the code.
\end{proposition}



\begin{example}
Consider $B_1=[\![4,2,2]\!]_2$ with stabilizer
generators $S_{B_1}=\{XXXX,ZZZZ\}$ and logical operators
$\overline{Z}_{1}=ZZII$, $\overline{X}_{1}=XIXI$,
$\overline{Z}_{2}=ZIZI$, $\overline{X}_{2}=XXII$. Then take the
subcodes $B_2=[\![4,1,2]\!]_2$ with stabilizer generators
$S_{B_1}\cup\{\overline{Z}_{1}\}$ and $B_3=[\![4,0,2]\!]_2$ with
stabilizer generators by $S_{B_1}\cup
\{\overline{Z}_{1},\overline{Z}_{2}\}$. Thus the coset code $[\![B_1/B_2]\!]$
has dimension $2$ with logical operators
$\{\overline{Z}_{1},\overline{X}_{1}\}$.  It will be used as the inner
code for the first level of concatenation.  Since
$B_3=\{|\mathbf{0}\rangle\}$, we have $[\![B_2/B_3]\!]\cong B_2$ with
logical operators $\{\overline{Z}_{2},\overline{X}_{2}\}$. It will be
used as the inner code on the second level of concatenation.

For the outer codes, take $A_1=[\![2,1,1]\!]_2$ with stabilizer
generators $S_{A_1}=\{ZZ\}$ and logical operators $\{ZI,XX\}$,
together with the trivial code $A_2=[\![2,2,1]\!]_2$ with logical
operators $\{ZI,XI,IZ,IX\}$. Then
$\overline{S}_{A_1}=\{\overline{Z}_{1}\overline{Z}_{1}\}$, and the
stabilizer $S_{\mathcal{C}}$ of the resulting GCQC is thus generated
by $S_I\cup\overline{S}_{A_1}$.  Furthermore, the set of logical
operators is given by $L_{\mathcal{C}}=\{
\overline{Z}_{1}I_4,\overline{X}_{1}\overline{X}_{1},
\overline{Z}_{2}I_4,\overline{X}_{2}I_4,
I_4\overline{Z}_{2},I_4\overline{X}_{2}\}$, where $I_4$ denotes the
identity operator on each of the $4$-qubit sub-blocks. The resulting
GCQC has parameters $\mathcal{C}=[\![8,3,2]\!]_2$.
\end{example}

\section{Parameters of GCQCs}
In order to derive the parameters of the GCQCs from our stabilizer
formalism, we will use the following lemma.  We keep the notation from
the previous sections. In addition, for a stabilizer code with
stabilizer generators $S$, we denote the normalizer group of $S$ by
$N(S)$.

\begin{lemma}

\label{lemma1}
Consider the restriction $\overline{W}_{\downarrow\bar{r}}$ and
$\overline{V}_{\downarrow\bar{s}}$ of any two elements $\overline{W}\in
N(\overline{S}_{A_i})$ and $\overline{V}\in N(\overline{S}_{A_j})$
($1\le i\le j\le m$) to sub-block $\bar{r}$ and $\bar{s}$
($r,s\in\{1,\ldots,N\}$), respectively, each block corresponding to
$n$ qubits obtained by mapping one coordinate of the outer code to the
$n$ qubits of the inner code. Then the product
$\overline{W}_{\downarrow\bar{r}}\cdot\overline{V}_{\downarrow{\bar{s}}}$ has weight at
least $d_i$, unless
$\overline{W}_{\downarrow\bar{r}}=\overline{V}_{\downarrow\bar{s}}=id$.
\end{lemma}

\begin{IEEEproof}
Case 1: $i=j$:\\
$\overline{W}_{\downarrow\bar{r}}\cdot\overline{V}_{\downarrow\bar{s}}$
is composed of the logical operators of $B_i$, whose distance is
$d_i$, thus
$\overline{W}_{\downarrow\bar{r}}\cdot\overline{V}_{\downarrow\bar{s}}$ has
weight at least $d_i$.

Case 2:
$i<j$:\\ $\overline{W}_{\downarrow\bar{r}}\cdot\overline{V}_{\downarrow\bar{s}}$
is composed of the logical operators from $B_i$ and $B_j$. Further,
$B_j\subset B_i$ implies $d_j>d_i$, thus
$\overline{W}_{\downarrow\bar{r}}\cdot\overline{V}_{\downarrow\bar{s}}$
has weight at least $d_i$.
\end{IEEEproof}

\begin{theorem}\label{theorem1}
Consider a GCQC
$\mathcal{C}=[\![\mathcal{N},\mathcal{K},\mathcal{D}]\!]_q$ which is
composed of $m$ outer codes $A_i=[\![N,K_i,D_i]\!]_{Q_i}$ and $m$ inner
codes $[\![B_{i}/B_{i+1}]\!]_q$ for $i=1,2,\dots,m$, where the code
$B_{i}=[\![n,k_{i},d_{i}]\!]_q$ is in the sub-code chain
$B_{m+1}\subset B_{m}\subset \dots \subset B_2\subset B_1$ and
$Q_i=q^{r_i}=q^{k_{i}-k_{i+1}}$. Let $A_\mu$ be the first degenerate
code regarding the ordering $A_1\succ A_2\succ\dots\succ A_m$ of the
outer codes.  Then the parameters of $\mathcal{C}$ are given as
\begin{enumerate}
\item
\begin{equation}
\label{eq:lenth}
\mathcal{N}=nN;
\end{equation}
\item
\begin{equation}
\label{eq:dimention}
\mathcal{K}=\sum_{i=1}^{m}(k_{i}-k_{i+1}){K_i};
\end{equation}
\item
\begin{equation}
\label{eq:distance1}
\mathcal{D}\ge\min\bigl\{d_1D_1, d_2D_2,\dots,d_{\mu-1}D_{\mu-1}, d_\mu\min_{\mu\le i\le m}\{D_i\}\bigr\}.
\end{equation}
\end{enumerate}
\end{theorem}
Note that if all outer codes are non-degenerate codes, it follows from
Eq.~(\ref{eq:distance1}) that
\begin{equation}
\label{eq:distance2}
\mathcal{D}\ge\min\{d_1D_1, d_2D_2,\dots,d_{m}D_{m}\}.
\end{equation}
If the first outer code a is degenerate code, then
\begin{equation}\label{eq:distance3}
\mathcal{D}\ge d_1\min_{1\le i\le m}\{D_i\}.
\end{equation}

\begin{IEEEproof}

1) Eq.~(\ref{eq:lenth}) is evidently true.

2) For each $A_i=[\![N,K_i,D_i]\!]_{Q_i}$, the number of independent
generators in $S_{A_i}$ is ${r_i}(N-K_i)$, which is also the number of
independent generators in $\overline{S}_{A_i}$. The number of
independent generators in $S_I$ is equal to $(n-k_1)N$. Therefore,
according to Proposition~\ref{pro:stabilizer}, we have
\begin{eqnarray}
\mathcal{K}&=&nN-(n-k_1)N-\sum_{i=1}^{m}{r_i}(N-K_i)\nonumber\\
&=&\sum_{i=1}^{m}(k_{i}-k_{i+1}){K_i},
\end{eqnarray}
where $r_i=k_{i}-k_{i+1}$ for $i=1,2,\dots,m$.

3) For a stabilizer code with stabilizer $S$, the minimum distance is
the minimum weight of an element in $N(S)\setminus S$. In other words,
it is the minimum weight of non-trivial logical operators. We consider
different cases how a logical operator of a QCQC can be composed
according to Proposition \ref{pro:stabilizer} (see
Fig.~{\ref{fig:stabilizerstructure}}).
\begin{figure}[htbp]
\centering
   \includegraphics[width=7cm,height=4cm]{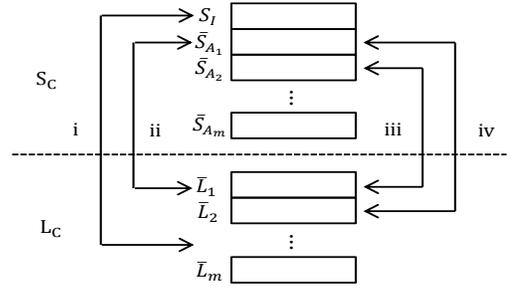}
\caption{Constitution of logical operators for a GCQC with all terms defined in the body.}
 \label{fig:stabilizerstructure}
\end{figure}

For the $i$th level of concatenation, we know that the distance of
$[\![B_i/B_{i+1}]\!]$ is at least $d_i$. As the distance of $A_i$ is
$D_i$, according to our replacement strategy, the non-trivial logical
operators obtained from $\overline{L}_{i}$ and $\overline{S}_{A_i}$
have weight at least $d_i$ on at least $D_i$ sub-blocks of length $n$.
Therefore, the minimal weight is at least $d_iD_i$.  Multiplying two
non-trivial elements $\overline{l}_i$ and $\overline{l}_j$ from two
different levels $i$ and $j$ with $i<j$, from Lemma~\ref{lemma1} the
product $\overline{l}_{ij}=\overline{l}_i\cdot \overline{l}_j$ must
have weight at least $d_i$ on at least $D_i$ sub-blocks of length $n$.
Denoting the weight of an operator $\overline{l}$ as ${\rm
  wgt}(\overline{l})$, for any element $\overline{l} \in
L_{\mathcal{C}}$ (see Eq.~(\ref{eq:logicalC})), we have
\begin{equation}
{\rm wgt}(\overline{l})\ge\min\{d_1D_1, d_2D_2,\dots,d_mD_m\}.
\end{equation}

Next we consider the minimal weight of the elements obtained by
multiplying a logical operator $\overline{l}\in L_{\mathcal{C}}$ by a
non-trivial stabilizer element $\overline{G}\in
S_{\mathcal{C}}$. First let $\bar{l}'=\overline{G}\cdot
\overline{l}_i$, where $\overline{G}\in S_I$ or $\overline{G}\in
\overline{S}_{A_j}$, and $\overline{l}_i\in \overline{L}_{i}$ for $1\le
i,j\le m$. Then we analyze ${\rm wgt}(\bar{l}')$ based on the
following cases (see Fig.~\ref{fig:stabilizerstructure}):
\renewcommand{\labelenumi}{(\roman{enumi})}
\begin{enumerate}
\item $\overline{G}\in S_I$: ${\rm wgt}(\bar{l}')\ge
  d_{i}D_{i}$ according to Eq.~(\ref{subsetsrelation}).
\item $\bar{G}\in \overline{S}_{A_j}$ and $i=j$: ${\rm
  wgt}(\bar{l}')\ge d_iD_i$.
\item $\bar{G}\in \overline{S}_{A_j}$ and $i<j$: ${\rm
  wgt}(\bar{l}')\ge d_iD_i$ according to
  Lemma~\ref{lemma1}.
\item $\overline{G}\in \overline{S}_{A_j}$ and $i>j$: ${\rm
  wgt}(\bar{l}')\ge d_j\times\max\{D_i,{\rm wgt}(G)\}$ according
  to Lemma~\ref{lemma1}. If $A_j$ is a non-degenerate outer code, then
  ${\rm wgt}(G)\ge D_j$, thus ${\rm wgt}(\bar{l}')\ge d_jD_j$. If $A_j$
  is a degenerate outer code, then there exist at least one
  non-trivial element $G\in S_{A_j}$ such that ${\rm
    wgt}(G)<D_j$, consider $D_i<D_j$ is probably true, thus ${\rm
    wgt}(\bar{l}')\ge d_j D_j$ is not guaranteed, but ${\rm
    wgt}(\bar{l}')\ge d_j D_i$ evidently is.
\end{enumerate}
Any non-trivial logical operator can be decomposed as a combination of
the four cases discussed above.
Now we are ready to get the distance
of a GCQC as shown by Eqs.~(\ref{eq:distance1}), (\ref{eq:distance2}),
and (\ref{eq:distance3}).
\end {IEEEproof}

Note that a degenerate outer code might also be viewed as a
non-degenerate code, but with a smaller distance.  As clarified by
Theorem~\ref{theorem1} and illustrated by the following example,
despite the larger minimum distance of the degenerate outer code, the
minimum distance of the resulting GCQC is not increased in general.
\begin{example}\label{Ex:2}
Let $B_1=[\![4,2,1]\!]_2$ with stabilizer generators
$S_{B_1}=\{ZZZZ,ZZII\}$ and logical operators
$\{\overline{Z}_{1}=XXXX,\overline{X}_{1}=IZZZ,
\overline{Z}_{2}=IIXX,\overline{X}_{2}=IZIZ\}$. The subcodes $B_1$ and
$B_2$ are $B_2=[\![4,1,2]\!]_2$ with stabilizer generators
$S_{B_1}\cup \{\overline{Z}_{1}\}$, and $B_3=[\![4,0,2]\!]_2$ with
stabilizer generators
$S_{B_1}\cup\{\overline{Z}_{1},\overline{Z}_{2}\}$.  Then
$[\![B_1/B_2]\!]$ of dimension $2$ with logical operators
$\{\overline{Z}_{B_1},\overline{X}_{B_1}\}$, is the inner code to be
used on the first level of concatenation, and $[\![B_2/B_3]\!]\cong B_2$
with logical operators $\{\overline{Z}_{2},\overline{X}_{2}\}$ is the
inner code to be used for the second level of concatenation.

The outer code $A_1=[\![5,1,2]\!]_2$ is a degenerate code with
stabilizer generators $S_{A_1}=\{XIIII,IXXXX,IZZZZ,IIIZZ\}$ and
logical operators $\{IZIZI,IXXII\}$. Furthermore, let
$A_2=[\![5,5,1]\!]_2$ be the trivial code with logical operators
$\{ZIIII,XIIII,IZIII,IXIII,\dots,IIIIZ,IIIIX\}$. According to our
replacement strategy, $\overline{G}=\overline{X}_{1}I_4I_4I_4I_4\in
\overline{S}_{A_1}$ and $\overline{l}=\overline{X}_{2}I_4I_4I_4I_4\in
\overline{L}_2$.  Note that the minimum weight of elements in
$\overline{L}_2$ is $2$, and ${\rm wgt}(\overline{l})=2$.  Now
consider the product $\overline{G}\cdot \overline{l}$ which is
obviously a logical operator of the resulting GCQC as well and which
plays the same role as $\overline{l}$.  It is easy to check that
multiplication by $\overline{G}$ reduces the weight of this logical
operator from $d_2D_2=2$ to $d_1D_2=1$, as predicted by
Eqs.~(\ref{eq:distance1}) and (\ref{eq:distance3}).

In fact, $A_1$ can also be viewed as a non-degenerate code with
parameters $[\![5,1,1]\!]_2$. This gives the lower bound ${\rm
  wgt}(G)\ge D_1$, and therefore ${\rm wgt}(\overline{G}\cdot
\overline{l})\ge d_1\times\max\{D_2,D_1\}\ge d_1D_1=1$, which is
consistent with Eq.~(\ref{eq:distance2}). In summary, the
resultant GCQC has parameters $[\![20,6,1]\!]_2$, but not
$[\![20,6,2]\!]_2$ as one would expect for non-degenerate codes.
\end{example}

\section{Discussion}
We have developed the structure of the stabilizer and logical
operators of generalized concatenated quantum codes. With the help of
quantum coset codes $[\![B_i/B_{i+1}]\!]$, the resulting code can be
considered as an abstract tensor product of codes $C_i$ corresponding
to the $i$th level of concatenation. For the code $C_i$, the lower
bound on the minimum distance is $d_iD_i$. This lower bound is met only
if all the non-identity entries of some logical operator of minimum
weight $D_i$ of $A_i$ are mapped onto the logical operators of minimum
weight $d_i$ of $[\![B_{i}/B_{i+1}]\!]$. In some cases, it is possible
to use a clever map to improve the minimum distance of $C_i$ and
thereby that of the resulting code.

\begin{example}
Take both $A_1$ and $B_1$ as $[\![2,1,1]\!]_2$ with stabilizer
generator $\{ZZ\}$ and logical operators
$\{\overline{Z}=ZI,\overline{X}=XX\}$. Take $B_2=\{|00\rangle\}$ as
the trivial one-dimensional code. Then $[\![B_1/B_2]\!]\cong B_1$ with
logical operators $\{\overline{Z}_{1}=ZI,\overline{X}_{1}=XX\}$. Now
we swap the role of the logical $X$- and the logical $Z$-operator of
$[\![B_1/B_2]\!]$. In other words, we let $\overline{Z}'_{1}=XX$,
$\overline{X}'_{1}=ZI$.  Then according to our replacement strategy,
we obtain a concatenated code $[\![4,1,2]\!]_2$ with stabilizer
generators $\{ZZII,IIZZ,XXXX\}$ and logical operators $\{XXII,ZIZI\}$,
while the original choice of logical operators for $B_1$ would only
give a code $[\![4,1,1]\!]_2$.
\end{example}
This example indicates that the minimum distance of the resulting GCQC
might be significantly improved compared to the lower bound when a
deliberate nesting strategy is used. That is because such a strategy
could be used to optimize the weight distribution for the logical
operators of the inner code $B_1$.  The stabilizer of the quantum
coset codes $[\![B_{i}/B_{i+1}]\!]$ depends on this choice, and
hence the parameters of the inner codes as well. In combination with
suitable chosen outer codes, the error-correcting capacity of
component codes could be exploited efficiently and the overall
performance might be better.





\section*{Acknowledgment}

YJW is supported by NSERC, AITF and RFDP.
The CQT is funded by the Singapore MoE and the NRF as part of the Research Centres of Excellence programme.
BZ and BCS are each supported by NSERC and CIFAR.

\end{document}